\documentclass[12pt]{article}
\def\theequation{\arabic{section}.\arabic{equation}}
\usepackage{amssymb}

\newcounter{rown}

\begin{document}
\renewcommand{\thefootnote}{\fnsymbol{footnote}}
\renewcommand{\theequation}{\thesection.\arabic{equation}}
\title{Irreps and Off-shell Invariant Actions of the $N$-extended Supersymmetric Quantum Mechanics}
\author{Francesco Toppan${}^{a}$\thanks{{\em e-mail: toppan@cbpf.br}}
\\ \\
${}^a${\it CBPF, Rua Dr.}
{\it Xavier Sigaud 150,}
 \\ {\it cep 22290-180, Rio de Janeiro (RJ), Brazil}}
\maketitle
\begin{abstract}
The complete classification of the irreducible representations of the $N$-extended
one-dimensional supersymmetry algebra linearly realized on a finite number of fields
is presented. Off-shell invariant actions of one-dimensional supersymmetric sigma models are
constructed. The role of both Clifford algebras and the Cayley-Dickson's doublings of algebras
in association with the $N$-extended supersymmetries is discussed. We prove in specific examples that the
octonionic structure constants enter the $N=8$ invariant actions as coupling constants. 
We further explain how to relate one-dimensional supersymmetric quantum mechanical systems to the
dimensional reduction of higher-dimensional supersymmetric theories. 
\end{abstract}
%\centerline{\copyright Francesco Toppan}
\vfill
\rightline{CBPF-NF-003/06}
{\em Talk given at the Fifth International Conference on Mathematical Methods in Physics --- IC2006,
		 April 24-28 2006,
		 CBPF, Rio de Janeiro, Brazil}

\newpage

\section{Introduction}

The classification of the irreducible representations of the one-dimensional $N$-extended supersymmetry has
been given in \cite{krt}. In this talk we briefly summarize the main results contained in that paper and discuss at length
one of the most promising frameworks for their application, based on the concept of dimensional reduction (or, conversely, ``oxidation" to higher dimensions \cite{oxi}). Many properties of higher-dimensional supersymmetric theories can be recovered in
terms of the associated one-dimensional supersymmetric quantum mechanical systems obtained through a dimensional
reduction procedure. In one dimension, powerful mathematical tools are available. They are based on Clifford
algebras, division algebras and, more generally, the Cayley-Dickson's doubling of an algebra. We prove that the division algebras structure constants not only enter the supersymmetry transformations, but they also
appear as coupling constants in off-shell invariant actions. The results here discussed open the way to the
construction of off-shell invariant actions for large values of $N$, when the superfield formalism is not available. Already in \cite{krt} we were able to 
produce an $N=8$ supersymmetric quantum mechanical system, not previously identified in the literature.
The Cayley-Dickson's connection with the extended supersymmetry can also shed light to a possible
off-shell formulation of the $11$-dimensional supergravity thought as low-energy limit of the $M$-theory.

\section{Grassmann algebras, Clifford algebras and the $N$-extended $D=1$
supersymmetry algebra}
The one-dimensional, $N$-extended supersymmetry algebra is a very fundamental
mathematical structure. This explains the reason why it can be applied to several
different mathematical problems ranging from Morse theory \cite{wit} to the index theorems.
Indeed, it is the simplest (this word should not be intended in technical term, because it is
not a simple superLie algebra) example of a graded algebra, admitting $N$ odd generators and
a single even generator, the central charge. Its mathematical importance becomes transparent
when we relate it to two other mathematical structures, the Grassmann algebra and the Clifford algebra. 
The Grassmann algebra is the enveloping algebra generated by the $N$ generators $\theta_a$ ($a=1,\ldots, N$)
satisfying the relation
\begin{eqnarray}
\theta_a\theta_b+\theta_b\theta_a &=& 0.
\end{eqnarray}
The Clifford algebra is the enveloping algebra generated by the $N$ generators $\gamma_i$ ($i=1,\ldots , N$)
satisfying the relation
\begin{eqnarray}
\gamma_i\gamma_j+\gamma_j\gamma_i &=& 2\eta_{ij} {\bf 1},
\end{eqnarray}
where $\eta_{ij}$ is a diagonal matrix and ${\bf 1}$ is the identity operator.
\par
In order to ``promote" the basic relation of the Clifford algebra as the constituent relation of a
(super)Lie algebra ${ G}$ (with a mapping ${ G}\times { G}\rightarrow { G}$) we have to interpret the 
$N$ $\gamma_i$ generators on the l.h.s. as the odd elements of a super-Lie algebra (we will call them $Q_i$'s) and we have to add the identity
${\bf 1}$ as an even element of the super-algebra. It corresponds to a central extension $z$ since ${\bf 1}$ commutes with
the $\gamma_i$'s. We are thus led to a superalgebra (the $N$-extended (pseudo) supersymmetry algebra) with total number of $N$ odd elements $Q_i$ and a single even
element, the central extension $z$.
The $N$-susy superalgebra is defined by the relations
\begin{eqnarray}\label{nsusyalgebra}
\{Q_i,Q_j\}&=& \eta_{ij} z,\nonumber\\
\relax [Q_i, z] &=& 0
\end{eqnarray}
If $\eta_{ij}\equiv\delta_{ij}$, the $N$-extended pseudo superalgebra is called the one-dimensional $N$ extended supersymmetry algebra
(from now on, for short, $N$-susy).\par
In physics, the central extension $z$ is denoted with ``$H$" and called the hamiltonian. The eigenspace
of a supersymmetric quantum mechanical problem is reduced to either a Grassmann algebra (for the vacuum states) or to a Clifford algebra. On the other hand, more information is needed to construct, e.g., off-shell
invariant actions of the supersymmetric quantum mechanics. The correct setting is furnished by the 
so-called $D$-module representations, i.e. linear transformations acting on multiplets of fields, bosonic and fermionic, depending on a single time-variable $t$. In the following we will classify the $D$-module
irreducible representations of the $N$-susy algebra and use them to construct new non-trivial off-shell invariant actions for $D=1$ $N$-extended sigma-models.

\section{The irreps of the $N$-extended $D=1$ supersymmetry revisited}

It is well-known that the Clifford algebras irreps can be classified in terms of division algebras \cite{{abs},{por},{oku}}.
On the other hand, the Clifford algebras are associated to the $D=1$ $N$-extended supersymmetry, as we have seen. The irreps of the (\ref{nsusyalgebra}) $N$-susy algebra are given by multiplets of fields, alternatively bosonic and fermionic, of different mass-dimensions, specified by the set of numbers $(n_1,n_2,n_3, \ldots, n_l)$, with $l\geq 2$ denoting the length of the multiplet. The total number of bosonic (fermionic) fields in the multiplet is given by $n$, with
$n=n_1+n_3+n_5+\ldots = n_2+n_4+\ldots$. 
The admissible multiplets, for a given $N$, are recovered from the ``root multiplets" of type $(n,n)$, which carry a representation of the $N$-susy algebra
expressed by the generators
\begin{eqnarray}\label{length2irrep}
Q_i&=& \frac{1}{\sqrt 2}\left( \begin{tabular}{cc} $0$& $\sigma_i$\\
${\widetilde \sigma}_i\cdot H$& $0$
\end{tabular}
\right) 
\end{eqnarray}
where the $\sigma_i$ and ${\widetilde\sigma}_i$ are matrices entering a Weyl type (i.e. block antidiagonal) 
irreducible representation of a $D$-dimensional (with $D=N$) Clifford algebra 
relation
\begin{eqnarray}\label{weylclifford}
\Gamma_i =\left( \begin{tabular}{cc} $0$& $\sigma_i$\\
${\widetilde \sigma}_i$& $0$
\end{tabular}
\right)\quad &,&\quad\{ \Gamma_i,\Gamma_j\}= 2\delta_{ij} 
\end{eqnarray}
The $Q_i$'s in (\ref{length2irrep}) are supermatrices with vanishing bosonic 
and non-vanishing fermionic blocks. The total number $2n$ of bosonic plus fermionic fields entering a multiplet
is given by the size of the corresponding gamma matrices. The remaining multiplets, for $l\geq 3$, are obtained
through a ``dressing procedure", see \cite{pt}, obtained by repeated applications of the transformations,
\begin{eqnarray}\label{dressing}
Q_i &\mapsto & {{\widehat Q}_i}^{(k)} = S^{(k)}Q_i {S^{(k)}}^{-1}
\end{eqnarray}
realized by diagonal matrices $S^{(k)}$'s ($k=1,\ldots, 2d$) with entries ${s^{(k)}}_{ij}$
given by
\begin{eqnarray}\label{entries}
{s^{(k)}}_{ij} &=& \delta_{ij}(1-\delta_{jk}+\delta_{jk}H)
\end{eqnarray}
The ``dressed" supersymmetric operators ${\widehat Q}_i$ have entries with integral powers of the
hamiltonian $H$. On the other hand, only the regular dressed operators, admitting no entries with poles
$\frac{1}{H}$, are genuine supersymmetry operators, linearly acting on a finite multiplet of bosonic and
fermionic fields. The problem of classifying the irreducible representations of the $N$-extended supersymmetry algebra is reduced to the problem of classifying the regular dressed operators. This problem has been solved
in \cite{krt}, by making use of specific properties of the Clifford algebras irreps. For any $N$, all length-$3$
multiplets of the type $(n-k, n, k)$ are an irrep of the $N$-susy. On the other hand, length-$4$ irreps
exist for $N=3,5,6,7$ and $N\geq 9$, while length-$5$ irreps are present starting from $N\geq 10$.\par
Up to $N=8$,
the list of length-$4$ irreps is, e.g., given by the multiplets
{{
{ {{\begin{eqnarray}&\label{N8length4}
\begin{tabular}{|l|c|}\hline
$N=3   $&$(1,3,3,1)$\\ \hline
$N=5   $&$(1,5,7,3), ~(3,7,5,1),~(1,6,7,2), ~(2,7,6,1),~(2,6,6,2), ~(1,7,7,1)$\\ \hline
$N=6   $&$(1,6,7,2),~(2,7,6,1),~(2,6,6,2), ~(1,7,7,1)$\\ \hline
$N=7   $&$(1,7,7,1)$\\  \hline
\end{tabular}&\nonumber\\
&&\end{eqnarray}}} }  
}}
Similarly, the $28$ length-$4$ irreducible multiplets of the $N=9$ susy algebra are given by
the set of numbers $(h, 16-k, 16-h,k)$, with $h,k$ constrained to satisfy $h+k\leq 8$.\par
In each multiplet, the set of fields of higher mass-dimension transform under supersymmery as time-derivatives.
They are called auxiliary fields and can be used to construct manifestly supersymmetric invariants.\par
Several other properties of the irreducible multiplets can be identified. For instance, it is possible to
prove that a dual multiplet specified by the ``reverse" numbers $(n_k, n_{k-1},\ldots , n_1)$ is an irreducible
multiplet iff $(n_1,n_2,\ldots, n_k)$ is an irrep.\par
Furthermore, tensoring irreps (and decomposing them into irreps) gives rise to the notion of ``fusion algebra" of the $N$-extended supersymmetry discussed in \cite{krt}. The fusion algebra contains information about the
off-shell invariants which are associated to the $N$-susy algebra.

\section{The oxidation program and the dimensionally reduced theories}

The supersymmetric quantum mechanics with large number $N$ of supersymmetries can be regarded as a framework
to investigate specific properties of higher-dimensional supersymmetric theories. The simplest way to see this is through the dimensional reduction, where all space-dimensions are frozen and the only dependence left
is in terms of a time-like coordinate. The usefulness of this procedure is due to the fact that in such a framework we can dispose of powerful mathematical tools (essentially based on the available classification of Clifford algebras, as previously discussed) which are not available in higher dimensions.\par
It should be remembered that a four-dimensional field theory with $N$ extended supersymmetries corresponds,
once dimensionally reduced to one-dimension, to a supersymmetric quantum mechanics with four times ($4N$) the number
of the original extended supersymmetries \cite{dire}. The most interesting case, in the context of the unification program, 
corresponds to the eleven-dimensional supergravity (the low-energy limit of the $M$-theory), which is reduced to
an $N=8$ four-dimensional theory and later to an $N=32$ one-dimensional supersymmetric quantum mechanical 
system.\par
  In this section we will discuss the dimensional reduction of supersymmetric theories from $D=4$ to $D=1$ in some specific examples . We will prove how certain $D=4$ problems can be reformulated in a $D=1$ language.\par
It is convenient to start with the dimensional analysis of the following theories:
\par
{\em i}) the free particle in one (time) dimension ($D=1$) and,\par
for the ordinary Minkowski space-time ($D=4$) the\par
{\em iia}) the scalar boson theory (with quartic potential $\frac{\lambda}{4!}\phi^4$),
\par
{\em iib}) the Yang-Mills theory and, finally,\par
{\em iic}) the gravity theory (expressed in the vierbein formalism).\par
We further make the dimensional analysis of the above three theories when dimensionally reduced ({\em\`a la Scherk}) 
to a one (time) dimensional $D=1$ quantum mechanical system.\par
In the following we will repeat the dimensional analysis for the supersymmetric version of these
theories.\par
{\em Case i}) - {\em the $D=1$ free particle}\par
It is described by a dimensionless action ${ S}$ given by
\begin{eqnarray}
{ S} &=& \frac{1}{m}\int dt {\dot \varphi}^2\end{eqnarray} 
The dot denotes, as usual, the time derivative.
The dimensionality of the time $t$ is the inverse of a mass; we can therefore set ($[t]=-1$). By assuming $\varphi$ being dimensionless ($[\varphi]=0$),
an overall constant (written as $\frac{1}{m}$) of mass dimension $-1$ has to be inserted to make ${ S}$ adimensional.
Summarizing, we have, for the above $D=1$ model,
\begin{eqnarray}
[t]_{D=1} &=& -1,\nonumber\\
\relax [\frac{\partial}{\partial t}]_{D=1} &=& 1,\nonumber\\
\relax [\varphi]_{D=1} &=& 0,\nonumber\\
\relax [m]_{D=1} &=& 0,\nonumber\\
\relax [{ S}]_{D=1}&=& 0.
\end{eqnarray}
The suffix $D=1$ has been added for later convenience, since the dimensional analysis corresponds to
a one-dimensional model. 
\par
{\em Case iia}) -{\em the $D=4$ scalar boson theory}\par
The action can be presented as 
\begin{eqnarray}
{ S} &=& \int d^4x\left(\frac{1}{2} \partial_\mu\Phi\partial^\mu\Phi -\frac{1}{2}M^2\Phi^2-\frac{1}{4!}\lambda\Phi^4\right),
\end{eqnarray}
An adimensional action ${ S}$ is obtained by setting, in mass dimension,
\begin{eqnarray}
[\Phi]_{D=4} &=& 1,\nonumber\\
\relax [\partial_\mu]_{D=4} &=& 1,\nonumber\\
\relax [M]_{D=4} &=& 1,\nonumber\\
\relax [\lambda]_{D=4}&=&0.
\end{eqnarray}
{\em Case iib}) -{\em the $D=4$ pure QED or Yang-Mills theories.}\par
The gauge-invariant action is given by
\begin{eqnarray}
{ S} &=& \frac{1}{e^2}\int d^4xTr\left( F_{\mu\nu}F^{\mu\nu}\right),
\end{eqnarray}
where the antisymmetric stress-energy tensor $F_{\mu\nu}$ is given by
\begin{eqnarray}
F_{\mu\nu}&=& [{ D}_\mu,{ D}_\nu],
\end{eqnarray}
with ${ D_\mu}$ the covariant derivative, expressed in terms of the gauge connection $A_\mu$
\begin{eqnarray}
{ D}_\mu &=&\partial_\mu - e A_\mu .
\end{eqnarray}
$e$ is the charge (the electric charge for QED). The action is adimensional, provided that
\begin{eqnarray}
[A_\mu]_{D=4} &=& 1,\nonumber\\
\relax [F_{\mu\nu}]_{D=4} &=& 2,\nonumber\\
\relax [e]_{D=4} &=& 0.
\end{eqnarray}
{\em iic}) -{\em The pure gravity case.}\par
The action is constructed, see \cite{wb} for details, in terms of the determinant ${ E}$ of the vierbein ${e_\mu}^a$
and the curvature scalar ${R}$. It is given by
\begin{eqnarray}
{ S} &=& \frac{-6}{8\pi G_N}{\int d^4x { E}{ R}}
\end{eqnarray}
The overall constant (essentially the inverse of the gravitational constant $G_N$) is now dimensional
($[G_N]_{D=4}= -2$). 
The adimensional action is recovered by setting
\begin{eqnarray}
\relax [{e_\mu}^a ]_{D=4} &=& 0,\nonumber\\
\relax [{ R}]_{D=4}&=& 2.
\end{eqnarray}
Let us now discuss the dimensional reduction from $D=4\Rightarrow D=1$. Let us suppose that the three space dimensions
belong to some compact manifold ${ M}$ (e.g. the three-sphere $S^3$) and let us freeze the dependence of the fields on the
space-dimensions (the application of the time derivative $\partial_0$ leads to non-vanishing results, while the application
of the space-derivatives $\partial_i$, for $i=1,2,3$, gives zero).
Our space-time is now given by ${\mathbb{R}}\times { M}$.
We get that the integration over the three space variables contributes just to an overall factor, the volume of the
three-dimensional manifold ${ M}$. Therefore
\begin{eqnarray}
\int d^4x &\equiv& Vol_{ M}\cdot \int dt 
\end{eqnarray} 
Since 
\begin{eqnarray}
[Vol_{ M}]_{D=4} &=& -3
\end{eqnarray}
we can express $Vol \equiv \frac{1}{m^3}$, where $m$ is a mass-term. A factor $\frac{1}{m}$ contributes as an
overall factor in the one-dimensional theory, while the remaining part $\frac{1}{m^2}$ can be used to rescale
the fields. We have, e.g., for the dimensional reduction of the scalar boson theory
that
\begin{eqnarray}
\varphi_{D=1}&\equiv& \frac{1}{m}\phi_{D=4}
\end{eqnarray}   
The dimensional reduction of the scalar boson theory {\em ii a}) is therefore given by
\begin{eqnarray}
{ S} &=& \frac{1}{m}\int dt \left( \frac{1}{2}{\dot\varphi}^2 -\frac{1}{2}M^2\varphi^2 +\lambda_{D=1}\frac{1}{4!}
\varphi^4\right)
\end{eqnarray}
where we have
\begin{eqnarray}
[\varphi]_{D=1} &=& 0,\nonumber\\
\relax [M]_{D=1}&=& 1,\nonumber\\
\relax [{\lambda_1}]_{D=1} &=& 2.
\end{eqnarray}
The $D=1$ coupling constant $\lambda_1$ is related to the $D=4$ adimensional coupling constant $\lambda$
by the relation
\begin{eqnarray}
\lambda_1 &=& \lambda m^2.
\end{eqnarray}
We proceed in a similar way in the case of the Yang-Mills theory. We can rescale the $D=4$ Yang-Mills fields $A_\mu$
to the $D=1$ fields $B_\mu = \frac{1}{m} A_\mu$. The $D=1$ charge $e$ is rescaled to $e_1 = e m$.
We have, symbolically, for the dimensional reduced action, a sum of terms of the type
\begin{eqnarray}
{ S} &=& \frac{1}{m} \int dt \left( {\dot B}^2 +e_1{\dot B} {B^2} + {e_1}^2 B^4\right)
\end{eqnarray}
where
\begin{eqnarray}
\relax [B]_{D=1} &=& 0,\nonumber\\
\relax [e_1]_{D=1} &=& 1.
\end{eqnarray}
The situation is different for what concerns the gravity theory. In that case the overall factor
${{Vol}_{ M}}/{G_N}$ produces the dimensionally correct $\frac{1}{m}$ overall factor of the one-dimensional
theory. This implies that we do not need to rescale the dimensionality of the vierbein ${e_\mu}^a$ and of the
curvature. Summarizing, we have the following results
\begin{eqnarray}&\label{tabledim}
\begin{tabular}{|cccrcr|}\hline
scalar boson & $\Phi$&:& $\relax [\Phi]_{D=4}=1$ &$\Rightarrow$  &$\relax [\Phi]_{D=1}=0$\\ \hline
gauge connection & $A_\mu$&:& $\relax [A_\mu]_{D=4}=1$ &$\Rightarrow$  &$\relax [A_\mu]_{D=1}=0$\\ \hline
vierbein & ${e_\mu}^a$&:& $\relax [{e_\mu}^a]_{D=4}=1$ &$\Rightarrow$  &$\relax [{e_\mu}^a]_{D=1}=0$\\ \hline
electric charge & $e$&:& $\relax [e]_{D=4}=0$ &$\Rightarrow$  &$\relax [e]_{D=1}=1$\\ \hline
\end{tabular}&\nonumber\\
&&\end{eqnarray}
Let us now discuss the $N=1$ supersymmetric version of the three $D=4$ theories above. 
We have at first the chiral multiplet, described in \cite{wb}, in terms of the chiral superfields $\Phi$, ${\overline\Phi}$. Next the vector multiplet $V$, the vector-multiplet in the Wess-Zumino gauge, the supergravity multiplet in terms of vierbein and gravitinos and, finally, the gauged supergravity multiplet
presenting an extra set of auxiliary fields. The total content of fields is given by the
following table, which presents also the $D=4$ and respectively the $D=1$ dimensionality of the
fields (in the latter case, after dimensional reduction). We have
\begin{eqnarray}&\label{tablesusy}
\begin{tabular}{|ccl|}\hline
chiral multiplet & :&$\Phi$, ${\overline \Phi}$\\ 
fields content & :&$(2,4,2)$\\
$D=4$ dimensionality & :&$\relax [1,\frac{3}{2}, 2]_{D=4}$ \\
$D=1$ dimensionality & :&$\relax [0,\frac{1}{2},1]_{D=1} $\\ \hline
vector multiplet & :&$V=V^\dagger$\\ 
fields content & :&$(1,4,6,4,1)$\\
$D=4$ dimensionality & :&$\relax [0,\frac{1}{2}, 1,\frac{3}{2}, 2]_{D=4}$ \\
$D=1$ dimensionality & :&$\relax [-1,-\frac{1}{2}, 0,\frac{1}{2},1]_{D=1} $\\ \hline
vector multiplet & :&$V$ in the WZ gauge\\ 
fields content & :&$(3,4,1)$\\
$D=4$ dimensionality & :&$\relax [1,\frac{3}{2}, 2]_{D=4}$ \\
$D=1$ dimensionality & :&$\relax [0,\frac{1}{2},1]_{D=1} $\\ \hline
supergravity multiplet & :&${e_\mu}^a$, ${{\psi}_\mu}^\alpha$\\ 
fields content & :&$(16,16)$\\
$D=4$ dimensionality & :&$\relax [0,\frac{1}{2}]_{D=4}$ \\
$D=1$ dimensionality & :&$\relax [0,\frac{1}{2}]_{D=1} $\\ \hline
gauged sugra multiplet & :&${e_\mu}^a$, ${{\psi}_\mu}^\alpha$, $b^i$\\ 
fields content & :&$(6,12,6)$\\
$D=4$ dimensionality & :&$\relax [0,\frac{1}{2}, 1]_{D=4}$ \\
$D=1$ dimensionality & :&$\relax [0,\frac{1}{2},1]_{D=1} $\\ \hline
\end{tabular}&\nonumber\\
&&\end{eqnarray}
Some comments are in order: the vector multiplet corresponds, in the $D=1$ language, to the
$N=4$ ``enveloping representation" \cite{krt} $(1,4,6,4,1)$. The latter is a reducible, but indecomposable representation
of the $N=4$ supersymmetry. Its irreducible multiplets are split into $(1,4,3,0,0)$ and $(0,0,3,4,1)$. 
The Wess-Zumino gauge, in the $D=1$ language, corresponds to select the latter $N=4$ irreducible multiplet,
whose fields present only non-negative dimensions.\par
The $N=2$ four-dimensional super-QED involves the coupling of a set of chiral superfields together with the vector
multiplet. Due to the dimensional analysis, the corresponding one-dimensional multiplet is the $(5,8,3)$ 
irrep of $N=8$ given by $(2,4,2)+(3,4,1)$.\par
For what concerns the supergravity theories, the original supergravity multiplet corresponds to four irreducible
$N=4$ one-dimensional multiplets, while the gauged supergravity multiplet is obtained, in the $D=1$ viewpoint,
in terms of three irreducible $N=4$ multiplets whose total number of fields is $(6,12,6)$.

\section{The off-shell invariant actions of the $N=4$ sigma models}

In the end of the eighties and the beginning of the nineties, the whole set of off-shell invariant actions
of the $N=4$ supersymmetries were produced \cite{{ikp}, one,onebis,two,three}, by making use of the superfield formalism. This result was reached
after slowly recognizing the multiplets carrying a representation of the one-dimensional $N=4$ supersymmetry.
The results discussed in this talk allows us to reconstruct, in a unified framework, all off-shell invariant
actions of the correct mass-dimension (the mass-dimension $d=2$ of the kinetic energy) for the whole set of the
$N=4$ irreducible multiplets. They are given by the $(4,4)$, $(3,4,1)$, $(2,4,2)$ and $(1,4,3)$ multiplets. 
\par
We are able to construct the invariants without using a superfield formalism. We use instead a construction which can be extended, how we will
prove later, even for large values of $N$, in the cases where the superfield formalism is not available.
We will use the fact that the supersymmetry generators $Q_i$'s act as graded Leibniz derivatives. Manifestly
invariant actions of the $N$-extended supersymmetry can be obtained by expressing them as
\begin{eqnarray}\label{adjointf}
{ I} &=& \int dt \left(Q_1\cdot \ldots \cdot Q_N f(x_1, x_2, \ldots , x_k)\right)
\end{eqnarray}
with the supersymmetry transformations applied to an arbitrary function of the $0$-dimensional fields $x_i$'s,
$i=1,\ldots, k$ entering an irreducible multiplet of the $N$-extended supersymmetry. Since the supersymmetry generators admit mass-dimension $=\frac{1}{2}$
(being the ``square roots" of the hamiltonian), we have that (\ref{adjointf}) is a manifestly supersymmetric
invariant whose lagrangian density $Q_1\ldots Q_Nf(x_1,\ldots, x_k) $ has a dimension $d=\frac{N}{2}$.
For $N=4$ the lagrangian density has the correct dimension of a kinetic term.\par
The $k$ variables $x_i$'s can be regarded as a coordinates of a $k$-dimensional manifold. The corresponding
actions can therefore be seen as $N=4$ supersymmetric one-dimensional sigma models evolving in a $k$-dimensional
target manifold. For each $N=4$ irrep we get the following results. In all cases below the arbitrary
$\alpha(x_i)$ function is given by $\alpha = \Box f(x_i)$. We get the following
list.\par
{\em i}) The $N=4$ $(4,4)$ case. We have:
\par
\begin{eqnarray}
Q_i(x,x_j;\psi,\psi_j)&=& (-\psi_i, \delta_{ij}\psi-\epsilon_{ijk}\psi_k; {\dot x}_i, -\delta_{ij}{\dot x} +\epsilon_{ijk}
{\dot x}_k)\nonumber\\
Q_4(x, x_j; \psi, \psi_j) &=& (\psi, \psi_j; {\dot x}, {\dot x}_j)
\end{eqnarray}
The most general invariant lagrangian ${ L}$ of dimension $d=2$ is given by
\begin{eqnarray}
\relax { L}&=& \alpha (\vec{x})[{\dot x}^2+{{\dot x}_j}^2-\psi{\dot\psi}-\psi_j{\dot\psi}_j]+\nonumber\\
&&\partial_x\alpha[\psi\psi_j{\dot x}_j-\frac{1}{2}\epsilon_{ijk}\psi_i\psi_j{\dot x}_k]+\nonumber\\
&&\partial_l\alpha[\psi_l\psi{\dot x}+\psi_l\psi_j{\dot x}_j+\frac{1}{2}\epsilon_{ljk}\psi_j\psi_k{\dot x}-\epsilon_{ljk}
\psi_j{\dot x}_k\psi]-\nonumber\\
&&-\Box\alpha\frac{1}{6}\epsilon_{ljk}\psi\psi_l\psi_k\psi_k
\end{eqnarray}
\par
~\par
{\em ii}) The $N=4$ $(3,4,1)$ case. We have:
 \par
 \begin{eqnarray}
Q_i(x_j;\psi,\psi_j; g)&=& (\delta_{ij}\psi-\epsilon_{ijk}\psi_k;{\dot x}_i;-\delta_{ij}g+\epsilon_{ijk}{\dot x}_k; 
-{\dot \psi}_i)\nonumber\\
Q_4(x_j; \psi, \psi_j; g_j) &=& (\psi_j; g, {\dot x}_j; {\dot \psi})\nonumber\\
\end{eqnarray}

The most general invariant lagrangian ${ L}$ of dimension $d=2$ is given by
\begin{eqnarray}
\relax { L}&=& \alpha (\vec{x})[{{\dot x}_j}^2+g^2-\psi{\dot\psi}-\psi_j{\dot\psi}_j]+\nonumber\\&&
\partial_i\alpha[\epsilon_{ijk}(\psi\psi_j{\dot x}_k+\frac{1}{2}g\psi_j\psi_k)-g\psi\psi_i+\psi_i\psi_j{\dot x}_j]-\nonumber\\
&&-
\frac{\Box\alpha}{6}\epsilon_{ijk}\psi\psi_i\psi_j\psi_k
\end{eqnarray}
\par
~\par
{\em iii}) The $N=4$ $(2,4,2)$ case. We have:
 \par
\begin{eqnarray}
Q_1(x,y;\psi_0,\psi_1,\psi_2,\psi_3; g,h)&=& (\psi_0,\psi_3;{\dot x},-g, h, -{\dot y}; -{\dot\psi}_1,{\dot\psi}_2)\nonumber\\
Q_2(x,y; \psi_0,\psi_1,\psi_2,\psi_3; g,h) &=& (\psi_3, \psi_0; {\dot y},-h,-g, {\dot x}; -{\dot \psi}_2,-{\dot\psi}_1)\nonumber\\
Q_3(x,y;\psi_0,\psi_1,\psi_2,\psi_3; g,h)&=& (-\psi_2,\psi_1; h, {\dot y} -{\dot x},-g;-{\dot\psi}_3,
{\dot\psi}_0)\nonumber\\
Q_4(x,y; \psi_0,\psi_1\psi_2,\psi_3; g,h) &=& (\psi_1, \psi_2; g, {\dot x}, {\dot y}, h;{\dot \psi}_0,{\dot\psi}_3)
\end{eqnarray}
The most general invariant lagrangian ${ L}$ of dimension $d=2$ is given by
\begin{eqnarray}
\relax { L}&=& \alpha (x,y)[{{\dot x}}^2+{\dot y}^2+g^2+h^2-\psi{\dot\psi}-\psi_j{\dot\psi}_j]+\nonumber\\&&
\partial_x\alpha[{\dot y}(\psi_1\psi_2-\psi_0\psi_3)+g(\psi_2\psi_3-\psi_0\psi_1)+h(\psi_1\psi_3+\psi_0\psi_2)]+\nonumber\\&&
\partial_y\alpha[-{\dot x}(\psi_1\psi_2-\psi_0\psi_3)-g(\psi_1\psi_3+\psi_0\psi_2)+h(\psi_2\psi_3-\psi_0\psi_1)]-\nonumber\\
&&-
{\Box\alpha}\psi_0\psi_1\psi_2\psi_3
\end{eqnarray}
\par
~\par
{\em iv}) The $N=4$ $(1,4,3)$ case. We have:
 \par
 \begin{eqnarray}
Q_i (x; \psi, \psi_j, g_j) &=& (-\psi_i; g_i, -\delta_{ij}{\dot x} +\epsilon_{ijk} g_k; \delta_{ij}{\dot\psi}-\epsilon_{ijk}
{\dot\psi}_k),\nonumber\\
Q_4 ( x; \psi, \psi_j; g_j) &=& (\psi; {\dot x}, g_j;{\dot \psi}_j)
\end{eqnarray}
The most general invariant lagrangian ${ L}$ of dimension $d=2$ is given by
\begin{eqnarray}\label{covariantN4action}
\relax{ L} &=& \alpha(x)[{\dot x}^2 -\psi{\dot\psi} -\psi_i{\dot\psi}_i +{g_i}^2]+\nonumber\\
&&
\alpha'(x)[ \psi g_i\psi_i -\frac{1}{2}\epsilon_{ijk}g_i\psi_j\psi_k] 
-\frac{\alpha''(x)}{6}[\epsilon_{ijk}\psi\psi_i\psi_j\psi_k]
\end{eqnarray}
It is worth recalling that $N=4$ is associated, as we have discussed, to the algebra of the quaternions.
This is why in the $(4,4)$, $(3,4,1)$ and $(1,4,3)$ cases the invariant actions can be written by making
use of the quaternionic tensors $\delta_{ij}$ and $\epsilon_{ijk}$. In the $(2,4,2)$ two fields
are dressed to be auxiliary fields and this spoils the quaternionic covariance property.  

\section{Octonions and the $N=8$ sigma-models}

Just as the $N=4$ supersymmetry is related with the algebra of the quaternions, the $N=8$ supersymmetry is related with the algebra of the octonions. More specifically, it can be proven that
the $N=8$ supersymmetry can be produced from
the lifting of the $Cl(0,7)$ Clifford algebra to $Cl(9,0)$. On the other hand, it is well-known \cite{crt2}, that the
seven $8\times 8$ antisymmetric gamma matrices of $Cl(0,7)$ can be recovered by the left-action of the imaginary
octonions on the octonionic space. As a result, the entries of the seven antisymmetric gamma-matrices of
$Cl(0,7)$ can be expressed in terms of the totally antisymmetric octonionic structure constants $C_{ijk}$'s.
The non-vanishing $C_{ijk}$'s are given by
\begin{eqnarray}
&C_{123}=C_{147}=C_{165}=C_{246}=C_{257}=C_{354}=C_{367}=1&.
\end{eqnarray}
The non-vanishing octonionic structure constants are associated with the seven lines of the Fano's projective plane,
the smallest example of a finite projective geometry, see \cite{bae}. 
The $N=8$ supersymmetry transformations of the various irreps can, as a consequence, be expressed in terms of the octonionic structure constants. This is in particular true for the dressed $(1,8,7)$ multiplet, admitting
seven fields which are ``dressed" to become auxiliary fields. It is an example of a multiplet
which preserves the octonionic structure since the seven dressed fields are related to the seven imaginary octonions. We have 
that the supersymmetry transformations are given by
\begin{eqnarray}
Q_i (x; \psi, \psi_j, g_j) &=& (-\psi_i; g_i, -\delta_{ij}{\dot x} +C_{ijk} g_k; \delta_{ij}{\dot\psi}-C_{ijk}
{\dot\psi}_k),\nonumber\\
Q_8 ( x; \psi, \psi_j; g_j) &=& (\psi; {\dot x}, g_j;{\dot \psi}_j)
\end{eqnarray}
for $i,j,k=1,\ldots, 7$.
The strategy to construct the most general $N=8$ off-shell invariant action of the $(1,8,7)$ multiplet
makes use of the octonionic covariantization principle. When restricted to an $N=4$ subalgebra, the invariant
actions should have the form of the $N=4$ $(1,4,3)$ action (\ref{covariantN4action}). This restriction can be done in
seven inequivalent ways (the seven lines of the Fano's plane).
The general $N=8$ action should be expressed in terms of the octonionic
structure constants. With respect to (\ref{covariantN4action}), an extra-term could in principle be present.
It is given by $\int dt \beta(x)C_{ijkl}\psi_i\psi_j\psi_k\psi_l$ and is constructed in terms of the octonionic tensor of
rank $4$ 
\begin{eqnarray}
C_{ijkl}&=&\frac{1}{6}\epsilon_{ijklmnp}C_{mnp}
\end{eqnarray}
(where $\epsilon_{ijklmnp}$ is the seven-indices totally antisymmetric tensor).
Please notice that the rank-$4$ tensor is obviously vanishing when restricting to the quaternionic subspace.
One immediately verifies that the term $\int dt \beta(x)C_{ijkl}\psi_i\psi_j\psi_k\psi_l$ breaks the $N=8$
supersymmetries and cannot enter the invariant action. For what concerns the other terms, starting from the general
action (with $i,j,k=1,\ldots, 7$)
\begin{eqnarray}\label{covariantN8action}
\relax{ S} &=& \int dt \{\alpha(x)[{\dot x}^2 -\psi{\dot\psi} -\psi_i{\dot\psi}_i +{g_i}^2]+\nonumber\\
&&
\alpha'(x)[ \psi g_i\psi_i -\frac{1}{2}C_{ijk}g_i\psi_j\psi_k] 
-\frac{\alpha''(x)}{6}[C_{ijk}\psi\psi_i\psi_j\psi_k]\}
\end{eqnarray}
we can prove that the invariance under the $Q_i$ generator ($=1,\ldots 7$) is broken by terms which, after integration by parts,
contain at least a second derivative $\alpha''$. We obtain, e.g., a non-vanishing term of the
type $\int dt \alpha''\frac{\psi}{2}C_{ijkl}g_j\psi_k\psi_l$. In order to guarantee the full $N=8$ invariance
(the invariance under $Q_8$ is automatically guaranteed)
we have therefore to set $\alpha''(x)=0$, leaving $\alpha$ a linear function in $x$.
As a result, the most general $N=8$ off-shell invariant action of the $(1,8,7)$ multiplet is given by
\begin{eqnarray}\label{N8invact}
{ S} &=& \int dt \{(ax +b) [{\dot x}^2 -\psi{\dot\psi} -\psi_i{\dot\psi}_i +{g_i}^2]+ a
[ \psi g_i\psi_i -\frac{1}{2}C_{ijk}g_i\psi_j\psi_k] \}
\end{eqnarray}
We can express this result in the following terms: the association of the $N=8$ supersymmetry with the octonions implies
that the octonionic structure constants enter as coupling constants in the $N=8$ invariant actions. The situation
w.r.t. the other $N=8$ multiplets is more complicated. The reason is due to the fact that the dressing of some
of the bosonic fields to auxiliary fields does not respect the octonionic covariance. The construction of the invariant actions
can however be performed along similar lines, the octonionic structure constants being replaced by the ``dressed"
structure constants. The procedure for a generic irrep is more involved than in the $(1,8,7)$ case. It is currently under
writing the full list of invariant actions for the $N=8$ irreps. The results will be reported elsewhere.
The method proposed is quite interesting because it allows in principle to construct the most general invariant actions.
It is worth mentioning that different groups, using $N=8$ superfield formalism, are still working in the problem of constructing the most general
invariant actions.

Let us close this section by pointing out that the only sign of the octonions is through their structure constants entering as
parameters in the (\ref{N8invact}) $N=8$ off-shell invariant action. (\ref{N8invact}) is an ordinary
action, in terms of ordinary associative bosonic and fermionic fields closing an ordinary $N=8$ supersymmetry algebra.

\section{The $N$-extended ``oxidized"' and ``Cayley-Dickson" supersymmetries}  

While the problem of constructing and classifying $N$-extended supersymmetric quantum mechanical systems can be formulated
for arbitrarily positive integer values of $N$, some special values of $N$ are more interesting than the other ones.
We will discuss here two series of such values, the so-called ``oxidized series" and the Cayley-Dickson series.
The first one is given by the values $N$ corresponding to the maximal number of extended supersymmetries which can be
linearly represented on a given set of bosonic and fermionic fields. Due to the Clifford algebra association, these
numbers can be easily seen to be given by $N=0,1,2,4~ mod~ 8$. The irreps of the non-oxidized supersymmetries can be recovered 
from the oxidized ones, making them worth to be studied separately. We get for the oxidized $N$'s the set of values
$N=1,2,4,8,9,10,12,16,17,\ldots $. The case $N=9$ is of particular interest. It is the first example of an extended supersymmetry beyond the $N=8$ barrier. We recall that no invariant off-shell action for sigma-models with a non-trivial metric has ever been constructed for $N>8$.  The irreducible multiplets of $N=9$ admit $16$ bosonic
and $16$ fermionic fields (they have been fully classified in \cite{krt}). In \cite{ber} it was constructed a $10$-dimensional supersymmetric YangMills theory presenting, after dimensional reduction to $D=1$, the $(9,16,7)$ multiplet of $N=9$.
Please notice that this is the maximal number of supersymmetries which can be linearly realized on this set of fields. A $D=10$
supersymmetric theory admits, from the one-dimensional viewpoint, $N=16$ supersymmetries. However, they are not necessarily
off-shell and linearly realized. A particular interesting subset of the oxidized supersymmetries is given by the
restriction $N=2^k$, therefore $N=1,2,4,8,16,32,\ldots$. We can call this series the Cayley-Dickson series. The first
four values indeed correspond to the four division algebra cases of the real, complex, quaternionic and octonionic
numbers. It is well-known that the Cayley-Dickson doubling, see e.g. \cite{kuztop} for explicit formulas, allows to construct each division
algebra by an iterated series of doubling of the real numbers. This construction does not stop to the octonions. The doubling
of the octonions produces an algebra, often named the algebra of the ``sedenions" \cite{lou}, which is no longer a division algebra
and does not possess a norm. The $15$ imaginary sedenions are, nevertheless, associated with the Clifford algebra
$Cl(0,15)$ and, consequently, to the $N=16$ supersymmetry. Such a supersymmetry is linearly realized on $128$ bosonic and $128$
fermionic fields. The eleven-dimensional supergravity is expressed in terms of the graviton with $44$ components,
the gravitinos ($128$ fermionic components) and a three form ($84$ bosonic components). In a $D=1$ viewpoint, the
$(44,128,84)$ multiplet carries a linear representation of the $N=16$ supersymmetry (an off-shell linear realization of the
$N=32$ supersymmetry implies $32,768$ bosonic and an equal number of fermionic fields). 
It is quite likely that the ``octonionic covariantization prescription" which has been advocated to produce $N=8$
off-shell invariant actions could be extended to the ``bi-octonions" expressing the $16$ bosonic fields carrying the
$N=9$ supersymmetry, as well as the sedenions leading to the $N=16$ supersymmetry. Just like the seven imaginary octonions are
accommodated in a triangle given by the Fano's plane, the $15$ imaginary sedenions are accommodated in a tetrahedron. 
Their non-vanishing structure constants are related to the set of $35$ lines connecting three imaginary sedenions. The Cayley-Dickson's doubling of the sedenions produces the $N=32$ supersymmetry. The control of such a large number of one-dimensional supersymmetries can provide the clue towards an off-shell formulation of the $11$-dimensional supergravity (the low-energy limit of the $M$-theory).

\section{Conclusions}

The supersymmetric quantum mechanics for large values of $N$, the number of extended supersymmetries, is a very active field
of research. Within the point of view here advocated, see also the discussions in \cite{glpr}, the supersymmetric quantum mechanics is investigated for its potential implications in the context of the
supersymmetric unifications of the interactions.
Along the years, much progress has been made to understand the nature of its irreducible representations and in applying them
to the construction of supersymmetric models. The role played by Clifford algebras was pointed out and used in
\cite{dr}, \cite{brr}, \cite{gr}.  This activity went parallel to the development of the superfield formalism based
on the notion of superspace, which can be
carried out up to $N=8$ (see \cite{abc} and references therein). The superspace is quite convenient for working with low
values of $N$. For larger values, however, the associated superfields are highly reducible and require the introduction
of constraints in order to extract the irreducible representations. In \cite{pt}, it was outlined the program of classifying
the irreducible representations, by making use of the connection between the $D=1$ Supersymmetry algebras and the associated Clifford algebras. This program was succesfully carried out in \cite{krt}. The main results have been reported
in this talk. The question of constructing and classifying $N$-extended supersymmetric quantum mechanical theories is still
an open problem, which has not been fully completed, even for $N=8$. For $N\geq 9$, it is not even known whether there
exists off-shell invariant sigma models with a non-trivial metric.\par 
We pointed out in this talk that the Cayley-Dickson's
doubling procedure produces an infinite series of algebras, ${\bf R}\rightarrow {\bf C}\rightarrow {\bf H}\rightarrow
{\bf O}\rightarrow \ldots $, whose structure constants enter the $N=1,2,4,8,16,\ldots$ supersymmetry transformations.
Up to $N=8$, we were further able to prove that the structure constants enter some off-shell supersymmetric invariant actions
as coupling constants. These results strongly suggest that both the sedenions (the Cayley-Dickson's doubling
of the octonionic algebra), and its double, could provide useful informations towards an off-shell formulation for the
$D=11$ supergravity. Indeed, the multiplet of the graviton in $D=11$ carries the $N=16$ linear supersymmetry transformations 
expressed by the structure constants of the sedenions. It is quite a technical challenge to investigate whether invariant
actions can be recovered as a consequence. \par Let us conclude this talk by pointing out that, in a different line, the representations' properties 
of the $N$-extended supersymmetric quantum mechanics are also being investigated in \cite{dfghil}.
\\{}~

\par {\large{\bf Acknowledgments}}{} ~\\{}~\par
It is a pleasure to acknowledge my collaborators Z. Kuznetsova and M. Rojas for useful discussions. This talk
is partly based on our common results. I thank E. Ivanov for interesting comments on the present status of the $N=8$
off-shell invariant actions.


\begin{thebibliography}{99}
\bibitem{krt} Z. Kuznetsova, M. Rojas and F. Toppan, JHEP 03 (2006) 098. 
\bibitem{oxi} H. Lu, C.N. Pope, E. Sezgin and K.S. Stelle, Nucl. Phys. {\bf B 456} 
(1995) 669; K.S. Stelle, ``Revising Supergravity and Super Yang-Mills
Renormalization" in ``{New Developments in Fundamental Interaction Theories}",
AIP 2001, eds. J. Lukierski and J. Rembieli\'nski, p. 108.
\bibitem{wit} E. Witten, Nucl. Phys. {\bf B 188} (1981) 513.
\bibitem{abs} M.F. Atiyah, R. Bott and A. Shapiro, Topology
(Suppl. 1) {\bf 3} (1964) 3.
\bibitem{por} I.R. Porteous, ``Clifford Algebras and the Classical Groups", 
Cambridge Un. Press, 1995.
\bibitem{oku} S. Okubo, J. Math. Phys. {\bf 32} (1991) 1657; {\em ibid.}
{\bf 32} (1991) 1669.
\bibitem{pt} A. Pashnev and F. Toppan, J. Math. Phys. {\bf 42} (2001) 5257.
\bibitem{dire} V. Rittenberg and S. Yankielowicz, Ann. Phys. {\bf 162} (1985) 273; M. Claudson and M.B. Halpern, Nucl. Phys. {\bf B 250} (1985) 689; R. Flume, Ann. Phys. {\bf 164} (1985) 189.
\bibitem{wb} J. Wess and J. Bagger, ``Supersymmetry and Supergravity", 2nd. ed., Princeton Un. Press (1992).
\bibitem{ikp} E.A. Ivanov, S.O. Krivonos and A.I. Pashnev, Class. Quantum Grav. {\bf 8} (1991) 19.
\bibitem{one} E.A. Ivanov, S.O. Krivonos and V. Leviant, J. Phys. {\bf A 22} (1989) 4201.
\bibitem{onebis} V.P. Berezovoj and A.I. Pashnev, preprint KFTI 91-20, Kharkhov (1991).
\bibitem{two} V.P. Berezovoj and A.I. Pashnev, Class. Quantum Grav. {\bf 13} (1996) 1699.
\bibitem{three} E.A. Ivanov and A.V. Smilga, Phys. Lett. {\bf B 257} (1991) 79; 
V.P. Berezovoj and A.I. Pashnev, Class. Quantum Grav. {\bf 8} (1991) 2141.
\bibitem{crt2} H.L. Carrion, M. Rojas and F. Toppan,
JHEP 0304 (2003) 040.
\bibitem{bae} J. Baez, Bull. Amer. Math. Soc. 39 (2002) 145.
\bibitem{ber} N. Berkovits, Phys. Lett. {\bf B 318} (1993) 104.
\bibitem{kuztop} Z. Kuznetsova and F. Toppan, hep-th/0610122.
\bibitem{lou} P. Lounesto, ``Clifford Algebras and Spinors", 2nd. ed., Cambridge Un. Press (2002).
\bibitem{glpr} S.J. Gates Jr., W.D. Linch and J. Phillips, hep-th/0211034; S.J. Gates Jr., W.D. Linch III, J. Phillips and L. Rana, Grav. Cosmol. {\bf 8} (2002) 96.
\bibitem{dr} M. de Crombrugghe and V. Rittenberg, Ann. Phys. {\bf 151} (1983) 99.
\bibitem{brr} M. Baake, M. Reinicke and V. Rittenberg, J. Math. Phys. {\bf 26} (1985) 1070.
\bibitem{gr} S.J. Gates Jr. and L. Rana, Phys. Lett. {\bf B 352} (1995) 50; {\em ibid.} {\bf B 369} (1996) 262.
\bibitem{abc} S. Bellucci, E. Ivanov, S. Krivonos and O. Lechtenfeld, Nucl. Phys. {\bf B 699} (2004) 226.
\bibitem{kra} M. Faux and S.J. Gates Jr., Phys. Rev. {\bf D 71} (2005) 065002.
\bibitem{dfghil} C.F. Doran, M.G. Faux, S.J. Gates Jr., T. Hubsch, K.M. Iga and G.D. Landweber, math-ph/05012016, 
math-ph/0603012, math-ph/0605269. 
\end{thebibliography}
\end{document}